\documentclass[reprint,
superscriptaddress,
nobibnotes,
amsmath,amssymb,
 aps,
prl,
showpacs
]{revtex4-1}

\usepackage{graphicx}\usepackage[dvipsnames]{xcolor}\usepackage{dcolumn}\usepackage{bm}
\usepackage[british]{babel}
\usepackage{subfigure}

\DeclareRobustCommand{\mean}[1]{\left\langle #1 \right\rangle}
\newcommand{\abs}[1]{\left| #1 \right|}

\definecolor{realnet}{rgb}{0.9,0.9,0.85}
\definecolor{unbiasnet}{HTML}{00BFC4}
\definecolor{biasnet}{HTML}{F8766D}

\begin{document}

\title{Generalized Hypergeometric Ensembles: \\ Statistical Hypothesis Testing in Complex Networks}

\author{Giona Casiraghi}
 \email{giona@ethz.ch}
\affiliation{ETH Z\"urich, Chair of System Design, Weinbergstrasse 56/58, 8092 Z\"urich, Switzerland
}
\author{Vahan Nanumyan}
 \email{vahan@ethz.ch}
\affiliation{ETH Z\"urich, Chair of System Design, Weinbergstrasse 56/58, 8092 Z\"urich, Switzerland
}
\author{Ingo Scholtes}
 \email{ischoltes@ethz.ch}
\affiliation{ETH Z\"urich, Chair of System Design, Weinbergstrasse 56/58, 8092 Z\"urich, Switzerland
}
\affiliation{AIFB, Karlsruhe Institute of Technology, Karlsruhe, Germany
}
 \author{Frank Schweitzer}\email{fschweitzer@ethz.ch}
\affiliation{ETH Z\"urich, Chair of System Design, Weinbergstrasse 56/58, 8092 Z\"urich, Switzerland
}

\date{5th August 2016}

\begin{abstract}
\emph{Statistical ensembles} of networks, i.e., probability spaces of all networks that are consistent with given aggregate statistics, have become instrumental in the analysis of complex networks.
Their numerical and analytical study provides the foundation for the inference of topological patterns, the definition of network-analytic measures, as well as for model selection and statistical hypothesis testing.
Contributing to the foundation of these data analysis techniques, in this Letter we introduce \emph{generalized hypergeometric ensembles}, a broad class of analytically tractable statistical ensembles of finite, directed and weighted networks.
This framework can be interpreted as a generalization of the classical configuration model, which is commonly used to randomly generate networks with a given degree sequence or distribution.
Our generalization rests on the introduction of \emph{dyadic link propensities}, which capture the \emph{degree-corrected} tendencies of pairs of nodes to form edges between each other.
Studying empirical and synthetic data, we show that our approach provides broad perspectives for model selection and statistical hypothesis testing in data on complex networks.
\end{abstract}

\maketitle

The analysis of data from the perspective of \emph{graphs} or \emph{networks} is a cornerstone in the study of complex systems.
Examples for network-based methods include (i) algorithms to detect communities of similar nodes, e.g., in social networks, (ii) statistical measures quantifying correlations like assortative mixing or clustering, and (iii) statistical techniques to infer significant patterns in biological or social systems.
Many of these methods are based on \emph{statistical ensembles}, probability spaces of networks with constraints imposed on characteristics like size, degree distribution, community structure, or motif statistics~\cite{Newman2001,Park2004,Karrer2011,Boccalettia2006,Garlaschelli2008,Bianconi2009,Fischer2015}.
The numerical and analytical study of such ensembles is of fundamental importance.
First, they serve as \emph{null models}, randomizing empirical networks to establish a \emph{baseline} of what is expected at random.
These are crucial to define quantitative measures~\cite{Newman2006,Newman2010}, to distinguish structural patterns from noise~\cite{Newman2006,Leicht2008,Karrer2011,Fischer2015}, and to study multi-layer~\cite{Menichetti2014,Wider2016} and temporal networks~\cite{Pfitzner2013_prl,Scholtes2014_natcomm,Karsai2012,Holme2015,Zhang2016}.
Moreover, they provide a foundation for model fitting and selection, with applications in community detection and hypothesis testing~\cite{Holland1981,Park2004,Newman2015,Peixoto2015x,Martin2016}.

Existing formulations of network ensembles generally fall into two classes.
A first class is based on \emph{generative models}, stochastic models that randomly generate networks satisfying given constraints.
Examples include classical models for random graphs with fixed number of nodes and edges~\cite{Erdos1960}, the configuration model generating networks with given degree sequence~\cite{Molloy1995}, or the stochastic block model generating topologies with known communities~\cite{Karrer2011}.
They can be used to define ensembles where (i) the characteristics of network realizations match those of an empirical network, and (ii) the probability of particular realizations is given by the underlying stochastic model.
While generative models provide a principled approach to define null models based on a well-defined random process, their application in practice is challenging.
Each generative model preserves only \emph{particular} characteristics, thus limiting its application as null model to specific scenarios.
Moreover, the analytical treatment of random processes generating networks, accurately satisfying given constraints, quickly becomes intractable~\footnote{We refer to analogies to \emph{kinetic theory}, cf. \cite{Park2004}}.
Thus, analytical solutions for such ensembles are scarce, instead relying on approximations obtained by generating a large number of samples.

A second class of ensembles addressing these limitations is the family of \emph{Exponential Random Graph Models} (ERGMs)~\cite{Holland1981}.
Different from generative models that preserve \emph{particular} quantities, ERGMs allow to fix expectations of \emph{arbitrary} characteristics observed in empirical networks.
Analogous to the grand canonical ensemble in equilibrium statistical mechanics~\cite{Park2004}, the exponential distribution naturally arises from the maximum entropy principle.
A benefit of ERGMs is that they provide a broad class of ensembles with tunable characteristics, making no assumptions about the underlying network generation process.
However, this generality also introduces problems.
Their analytical treatment requires a closed-form solution of the \emph{partition function}, which can generally be derived only for infinite networks with statistically independent links~\cite{Holland1981,Park2004}.
At the same time, sampling-based approximations are complicated by computational requirements and phenomena that can break ergodicity~\cite{Park2004,Foster2010,Fischer2015}.
Finally, the lack of knowledge about network formation processes, expressed in the choice of the maximum entropy distribution, poses problems in situations where we are interested in null models that capture specific random network generation processes.
These issues limit our ability to study patterns in large, but finite networks.

Extending the foundation of ensemble-based model selection and hypothesis testing, in this Letter we introduce an \emph{ab initio} alternative to the grand canonical ensemble formulation of complex networks.
The resulting class of \emph{generalized hypergeometric ensembles} is based on a simple generative model of complex networks.
It provides an analytically tractable generalization of the configuration model~\cite{Molloy1995} for directed, weighted networks with a fixed, finite number of nodes and edges, constrained by a given sequence of \emph{expected} degrees.
Incorporating \emph{dyadic propensities}, it further allows to encode arbitrary \emph{degree-corrected} tendencies of pairs of nodes to form edges between each other.
We show that this class of ensembles provides a powerful framework for model selection in complex networks.
Furthermore, we demonstrate (i) how hypotheses about topological patterns in networks can be tested statistically, and (ii) that this provides a new approach to test the statistical significance of community structures.

We introduce our methodology step by step.
Let us consider an empirical network consisting of repeated dyadic relations $(i,j)$ between nodes $i$ and $j$.
Such relational data can be represented as a \emph{multi-edge}, or \emph{weighted}, network $\hat{G}=(V,E)$, with a set $V$ of $n$ nodes, and a multiset $E \subseteq V \times V$ of (directed) edges.
We further define an adjacency matrix $\hat{\mathbf{A}}$, where entries $\hat{A}_{ij}\in\mathbb N_{0}$ capture the \emph{weight} of edge $(i,j)\in V \times V$, i.e. the multiplicity of edge $(i,j)$ in the multiset $E$.
For each node $i \in V$ we can define the (weighted) in- and out-degree as $\hat{k}_{\mathrm{in}}(i) := \sum_{j \in V} \hat{A}_{ji}$ and $\hat{k}_{\mathrm{out}}(i) := \sum_{j \in V} \hat{A}_{ij}$ respectively.
For undirected networks, the adjacency matrix is symmetric and $\hat{k}_{\mathrm{in}}(i)=\hat{k}_{\mathrm{out}}(i)=:\hat{k}_{i}$.
By definition, for the total number of multi-edges $m:=|E|$ we have $m=\sum_{i,j \in V}\hat{A}_{ij}=\sum_{i \in V}\hat{k}_{\mathrm{out}}(i) = \sum_{i \in V}\hat{k}_{\mathrm{in}}(i)$.

Our construction of a statistical ensemble follows the idea of the Molloy-Reed configuration model~\cite{Molloy1995}, which is to randomly shuffle the topology of a network $G$ while preserving node degrees.
The configuration model uses a \emph{node-centric sampling} approach, generating edges between randomly sampled pairs of nodes such that the \emph{exact} observed degrees of nodes are preserved.
Different from this, we utilize an \emph{edge-centric sampling} of $m$ edges from the set of all possible edges such that the sequence of \emph{expected} degrees of nodes is preserved.
For each pair of nodes $i,j$, we first define the maximum possible number $\Xi_{ij}$ of multi-edges that can exist between nodes $i$ and $j$ as $\Xi_{ij} := \hat{k}_{\mathrm{out}}(i) \hat{k}_{\mathrm{in}}(j)$ (cf. \cite{Newman2006,Karrer2011}), which can be conveniently represented in matrix form as $\mathbf{\Xi} := \left(\Xi_{ij}\right)_{i,j \in  V}$ for all pairs of nodes.

We can hence define a statistical ensemble based on the following generative model.
For each pair of nodes $i,j$, we sample edges from a set of $\Xi_{ij}$ possible multi-edges uniformly at random.
This can be viewed as an \emph{urn problem}~\cite{jacod2003probability} where edges to be sampled are represented by balls in an urn.
We specifically obtain an urn with $M=\sum_{i,j}\Xi_{ij}$ balls having $n^{2}=\abs{V\times V}$ different colours, representing all possible edges between a given pair of nodes.
The sampling of a network corresponds then to drawing exactly $m$ balls from this urn.
Each adjacency matrix $\mathbf{A}$, with entries $A_{ij}$ such that $\sum_{i,j} A_{ij}=m$, corresponds to one particular realization drawn from this ensemble.
The probability to draw exactly $\mathbf{A}=\{A_{ij}\}_{i,j \in  V}$ edges between each pair of nodes is given by the \emph{multivariate} hypergeometric distribution\footnote{We do not distinguish between the $n\times n$ adjacency matrix $\mathbf{A}$ and the $n^{2}\times 1$ vector obtained by stacking.}
\begin{equation}
	\label{eq:hypergeometricNet}
  \Pr(\mathbf{A}) = \dbinom{M}{m}^{-1} \prod_{i,j}\dbinom{\Xi_{ij}}{A_{ij}}.
\end{equation}
which provides an analytical expression for the probability of the given corresponding network $\hat{G}$.

For each pair of nodes $i,j \in V$, the probability to draw exactly $\hat{A}_{ij}$ edges between $i$ and $j$ is given by the marginal distributions of the multivariate hypergeometric distribution
\begin{equation}
	\label{eq:hypergeometricEdge}
  \Pr(A_{ij}=\hat{A}_{ij}) =  \dbinom{M}{m}^{-1} \dbinom{\Xi_{ij}}{\hat{A}_{ij}}\dbinom{M-\Xi_{ij}}{m-\hat{A}_{ij}}.
\end{equation}
For each pair of nodes $i,j$ we can further calculate the expected number of edges as $\langle A_{ij} \rangle = m \frac{\Xi_{ij}}{M}$.
Moreover, we can calculate the expected (weighted) in-degrees of all nodes by summing the columns (rows) of matrix $\langle A_{ij} \rangle$ (analogously for out-degrees):
\begin{equation}
\label{eq:hypergeometricDegree}
\langle k_{\mathrm{in}}(j) \rangle = \sum_{i \in V}{\langle A_{ij} \rangle} = m \frac{\sum_{i \in V} \hat{k}_{\mathrm{out}}(i) \hat{k}_{\mathrm{in}}(j)}{M} = \hat{k}_{\mathrm{in}}(j).\end{equation}

Eq.~\ref{eq:hypergeometricDegree} confirms that the \emph{expected} in- and out-degree sequence of realizations drawn from the resulting statistical ensemble corresponds to the degree sequence of the given network $\hat{G}$.
We thus arrive at a \emph{hypergeometric statistical ensemble}, which (i) provides a generalization of the configuration model for directed, multi-edge networks, (ii) has a fixed sequence of \emph{expected} degrees, and (iii) is analytically tractable for directed and undirected networks with and without self-loops and multiple edges.
Furthermore, we obtain a framework for the generalization of other generative models like, e.g., the multi-edge version of the Erd\"os-R\'enyi model~\cite{erdds1959random}, where only $n$ and $m$ are fixed, while there are no constraints on the degree sequence~\footnote{It corresponds to $\mathbf{\Xi}$ with $\Xi_{ij}=m^{2}/n^{2}=\;$const which directly results from $\mean{k_{\mathrm{in}}(i)}=\mean{k_{\mathrm{out}}(i)}=m/n$}.

The sampling procedure outlined above provides a parsimonious stochastic model for weighted, directed networks in which (i) the expected in- and out-degrees of nodes are fixed, and (ii) edges between these nodes are generated at random,
This stochastic model serves as a \emph{null model} because it considers only \emph{combinatorial effects} and \emph{no} additional correlations.
The question is to what extent the patterns in a given empirical network exhibit statistically significant \emph{deviations} from this null model.

To answer this question, we generalize the \emph{hypergeometric ensemble} as follows.
We introduce a matrix $\mathbf{\Omega}$ whose entries $\Omega_{ij}$ capture relative \emph{dyadic propensities}, i.e., the tendency of a node $i$ to form an edge \emph{specifically} to node $j$.
In particular, we assume that an entry $\Omega_{ij}$ captures the propensity that goes \emph{beyond} the tendency of a node $i$ to connect to a node $j$ that results from combinatorial effects, i.e., a \emph{degree-corrected preference} of $i$ linking to $j$ which accounts for the in-degree of $j$ and the out-degree of $i$.
The key idea of our generalized ensemble is to use the dyadic propensities $\Omega_{ij}$ to \emph{bias} the edge sampling process described above.
In analogy to the urn model, a biased sampling implies that the probability of drawing balls of a given color does not only depend on their number but also on the respective relative propensities.
The probability distribution resulting from such a biased sampling process is given by the multivariate \emph{Wallenius' non-central hypergeometric distribution}~\cite{Wallenius1963,Fog2008a,chesson1976non}:
\begin{equation}
	\label{eq:walleniusNet}
	\Pr(\mathbf{A})=\left[\prod_{i,j}{\dbinom{\Xi_{ij}}{A_{ij}}}\right]
         \int_{0}^{1}{\prod_{i,j}{\left(1-z^{\frac{\Omega_{ij}}{S_{\mathbf{\Omega}} }}\right)^{A_{ij}}}dz}
\end{equation}
with $S_{\mathbf{\Omega}}= \sum_{i,j} \Omega_{ij}(\Xi_{ij}-A_{ij})$.

Similar to the un-biased sampling, the probability to observe a particular number $\hat{A}_{ij}$ of edges between a pair of nodes $i$ and $j$ can again be calculated from the marginal distribution:
\begin{equation}
	\label{eq:walleniusEdge}
  \begin{aligned}
	\Pr(&A_{ij}=\hat{A}_{ij}) =  \dbinom{\Xi_{ij}}{\hat{A}_{ij}}\dbinom{M-\Xi_{ij}}{m-\hat{A}_{ij}}\cdot  \int_{0}^{1} \left[ \vphantom{e^\frac12}\right. \\
          &\left(
                    1 - z^{ \frac{\Omega_{ij}}{S_{\mathbf{\Omega}}}}
\right)^{\hat{A}_{ij}} \left(
                    1-z^{ \frac{\bar{\Omega}_{\setminus(i,j)}}{S_{\mathbf{\Omega}}}}
\right)^{m-\hat{A}_{ij}}
\left.\vphantom{\vphantom{e^\frac12}}\right]dz
  \end{aligned}
\end{equation}
where $\bar{\Omega}_{\setminus(i,j)} = (M-\Xi_{ij})^{-1}\sum_{(l,m)\in V\times V\backslash(i,j)}{\Xi_{lm}\Omega_{lm}}$.

The entries of the expected adjacency matrix \(\langle A_{ij}\rangle\) can be obtained by solving the system of equations described in~\cite{Fog2008a}.
Note that for the special case of a uniform dyadic propensity matrix $\mathbf{\Omega} \equiv \text{const}$, which corresponds to an unbiased sampling of edges, the integral in Eq.~\ref{eq:walleniusEdge} becomes $\binom{M}{m}^{-1}$ and we thus recover Eq.~\ref{eq:hypergeometricNet}.

A major advantage of the formalism outlined above is that, by specifying different dyadic propensities matrices $\mathbf{\Omega}$, we obtain a broad class of \emph{generalized hypergeometric ensembles}.
This allows us to encode a wide range of dyadic patterns in networks, while still obtaining an analytically tractable statistical ensemble from a simple and well-defined generative model.
In the following, we show how the class of generalized hypergeometric ensembles can be used for \emph{model selection} and \emph{hypothesis testing} in complex networks.

Starting with the first, we introduce a method to select which of the ensembles defined by different $\mathbf{\Omega}$ is the most plausible model for a given empirical network $\hat{G}$.
We recall that an (unbiased) random generation of edges between nodes with fixed (expected) in- and out-degrees corresponds to the generalized configuration model with $\mathbf{\Omega} \equiv \text{const}$.
Different models for the patterns present in the topology of a network $\hat{G}$ can be encoded in terms of \emph{different dyadic propensities} $\mathbf{\Omega}$.
Better models correspond to ensembles statistically closer to the observed network.
The \emph{statistical distance} between $\hat{G}$ and the ensemble identified by $\mathbf\Omega_{r}$ can be assessed by the \emph{Mahalanobis distance}~\cite{Mahalanobis1936}. This multivariate generalization of the $Z$-score captures how many standard deviations an observation is away (in the corresponding direction) from the expectation.
For each $\mathbf{\Omega}_{r}$, this provides a simple method to compare an empirical network $\hat{G}$ with adjacency matrix $\mathbf{\hat{A}}$ to the expected adjacency matrix $\langle \mathbf{A}\rangle_{r}$ (with associated covariance matrix $\mathbf{\Sigma}_{r}$).
The square of the Mahalanobis distance is
\begin{equation}\label{eq:1}
  D^{2}_{r}(\mathbf{\hat{A}}) = \left( \mathbf{\hat{A}} - \langle \mathbf{A} \rangle_{r}\right)^T \mathbf{\Sigma}_{r}^{-1} \left(\mathbf{\hat{A}} - \langle \mathbf{A} \rangle_{r} \right).
\end{equation}
For the distribution given by Eq.~\ref{eq:walleniusNet}, ${\langle \mathbf{A} \rangle}_r$ can be calculated analytically and its covariance matrix $\mathbf{\Sigma}_{r}$ can be approximated numerically~\cite{Fog2008}.
From a set of candidate models, the one with the smallest Mahalanobis distance is to be preferred, since the corresponding ensemble is statistically closest to an observed empirical network.
It thus best captures the topological patterns present therein.

In the following, we illustrate the resulting \emph{model selection} procedure using a synthetic toy example.
We generate observations according to the small-world lattice model proposed in~\cite{Watts1998}.
Starting from a ring lattice topology with $n=30$ nodes and $m=750$ directed multi-edges where each node is connected to its $K=5$ nearest neighbours via $L=5$ multi-edges each, we randomly rewire all edges with probability $\beta\in [0,1]$ retaining the in- and out-degree sequences.
This procedure produces synthetic network realizations $\hat G^\beta$ with adjacency matrices $\mathbf{\hat  A}^\beta$
that exhibit ring patterns with varying intensity -- from the perfect ring lattice when $\beta=0$ to a completely random network when $\beta=1$.
To demonstrate the \emph{model selection} procedure let us, without loss of generality, consider two candidate models:
the configuration model which generates no ring pattern, and a model which generates the correct ring pattern with varying strength.
Both can be encoded within the class of generalized hypergeometric ensembles in terms of dyadic propensities
\begin{equation}
   {\Omega}^c_{ij}  =
 \begin{cases}
  1 &  \text{for}\; j\in[i+1\mod{n},i+K\mod{n}],\\
  c &  \mathrm{otherwise, with}\; c \in [0,1].
  \end{cases}
\end{equation}
The parameter $c$ defines the intensity of the pattern, such that $c=0$ corresponds to perfect rings, while $c=1$ corresponds to the configuration model with no pattern.

\begin{figure}[ht] \centering
  \includegraphics[width=\columnwidth]{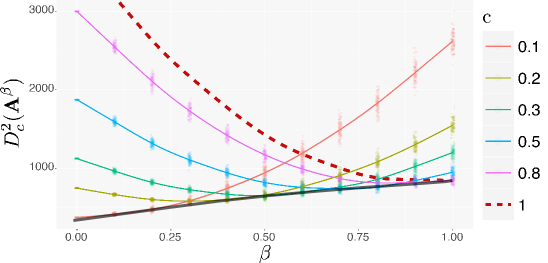}
  \caption{Mahalanobis distance $D^{2}_{c}(\mathbf{\hat A}^{\beta})$ for multiedge Watts-Strogatz networks with varying parameter $\beta \in \{0, 0.1, \ldots 1\}$, with $100$ realizations for each.  The dashed curve corresponds to configuration model.  Colored curves represent the average interpolated distance for the model with a ring pattern, with five different parameters $c \in [0.1, 0.8]$.  The black curve is the average distance for the best fitting ring model.
  \label{fig:strogatz}}
\end{figure}

For the observed network $\mathbf{\hat A}^\beta$ we compute its Mahalanobis distance from the configuration model ($\mathbf\Omega\equiv 1$).
The dashed line in Fig.~\ref{fig:strogatz} shows the results for networks obtained varying $\beta$.
As expected, the higher the rewiring probability $\beta$, the smaller the distance for the configuration model.
For each $\mathbf{\hat A}^\beta$ we compute the distance $D^{2}_{c}(\mathbf{\hat A}^{\beta})$ for different $c$, as shown in Fig.~\ref{fig:strogatz}.
Minimizing $D^{2}_{c}(\mathbf{\hat A}^{\beta})$ over $c$ corresponds to \emph{fitting} the model to the given network, using Mahalanobis distance as goodness-of-fit measure.
Fig.~\ref{fig:strogatz} shows that, as expected, the fitted $c$ grows with $\beta$.
We now compare the best fit of the ring model ($c<1$) to the configuration model ($c=1$).
As expected, the fitted ring model is the best model for all $\beta<1$, while for $\beta=1$ the two models coincide.

Above, we demonstrated how our ensemble formulation can be used to select (and fit) candidate models for complex networks.
Each of these models corresponds to a \emph{hypothesis} about topological patterns present in an empirical network.
Beyond the \emph{relative} comparison of candidate models, a major contribution of our framework is the ability to \emph{test hypotheses} about patterns in complex networks.
Consider an empirical network $\hat{G}$ with adjacency matrix $\mathbf{\hat{A}}$.
A hypothesis $H_0$ about the network (e.g. the presence of a ring pattern or community structures) can be encoded in terms of edge propensities $\mathbf{\Omega}_{0}$.
We then need to compute a $p$-value, i.e., the probability to draw a random realization that is, compared to the observed network, more extreme with respect to a test statistic.
For the multivariate distribution in Eq. \eqref{eq:walleniusNet}, the Mahalanobis distance (cf. Eq. \eqref{eq:1}) is a suitable statistic.
Therefore, a $p$-value for $H_0$ can be given in terms of the complementary cumulative distribution $\Pr\left[D^{2}_{0}(\mathbf{A})\geq D^{2}_{0}(\mathbf{\hat A})\right]$ where $\mathbf{A}$ is a random realization drawn from the generalized hypergeometric ensemble defined by $\mathbf{\Omega}_{0}$.
Under certain conditions~\footnote{$\Pr\left[D_{0}^{2}(\mathbf{A})\geq x)\right]$  converges to a $\chi^{2}$ distribution with $n^{2}-1$ degrees of freedom if $ \Xi_{ij} \gg m \text{ and } \Omega_{ij}=c ~ \forall i,j \in V$.}, there are closed-form expressions for $\Pr\left[D_{0}^{2}(\mathbf{A})\geq x\right]$.
However, in the following we resort to a sampling procedure, which is facilitated by the simplicity of the underlying generative model.

\begin{figure}[htbp] \centering
  \subfigure[\label{fig:karate:empirical}]{\includegraphics[width=.3\columnwidth]{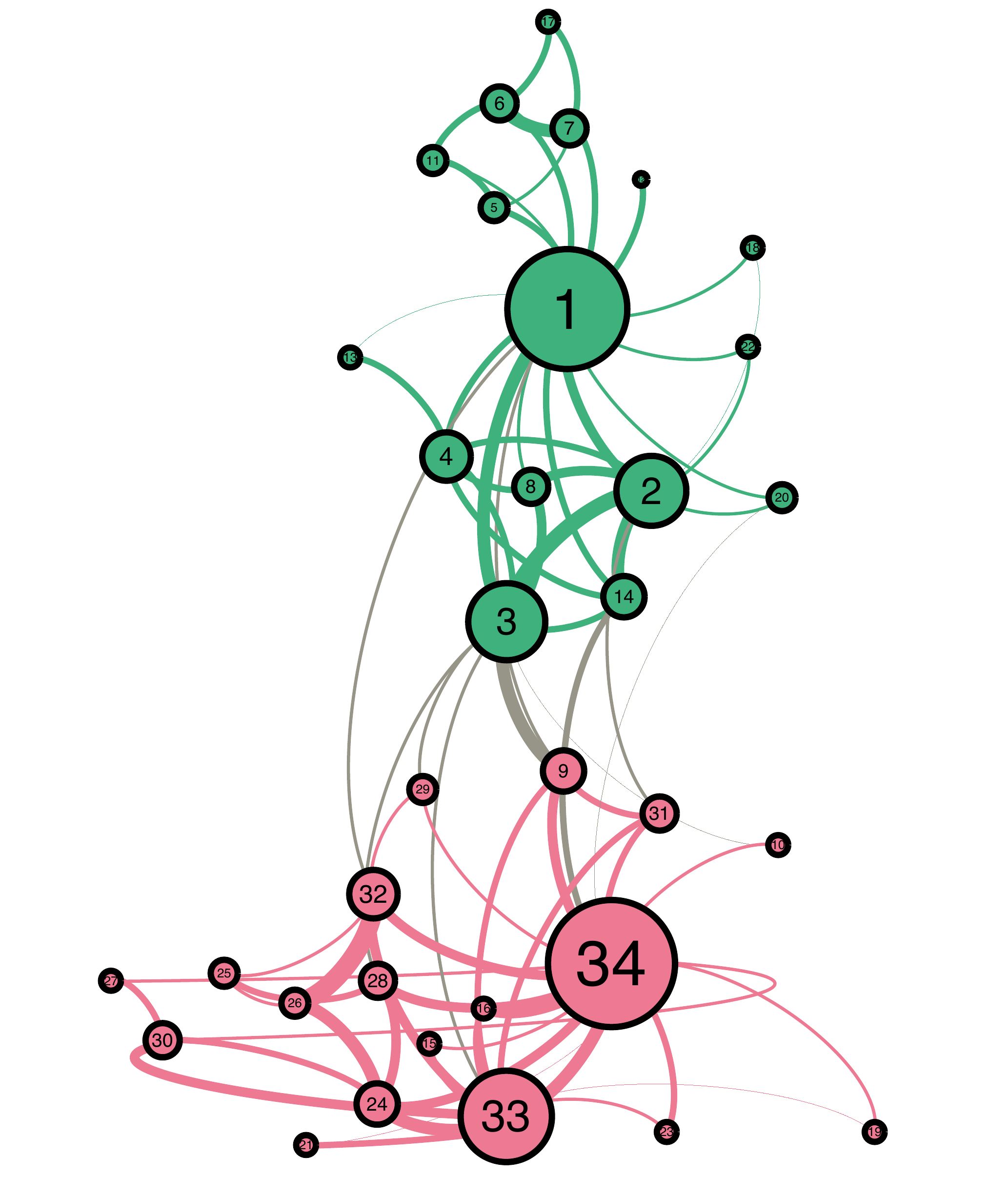}}\hfill
  \subfigure[\label{fig:karate:unbiased}]{\fcolorbox{unbiasnet}{unbiasnet}{\includegraphics[width=.3\columnwidth]{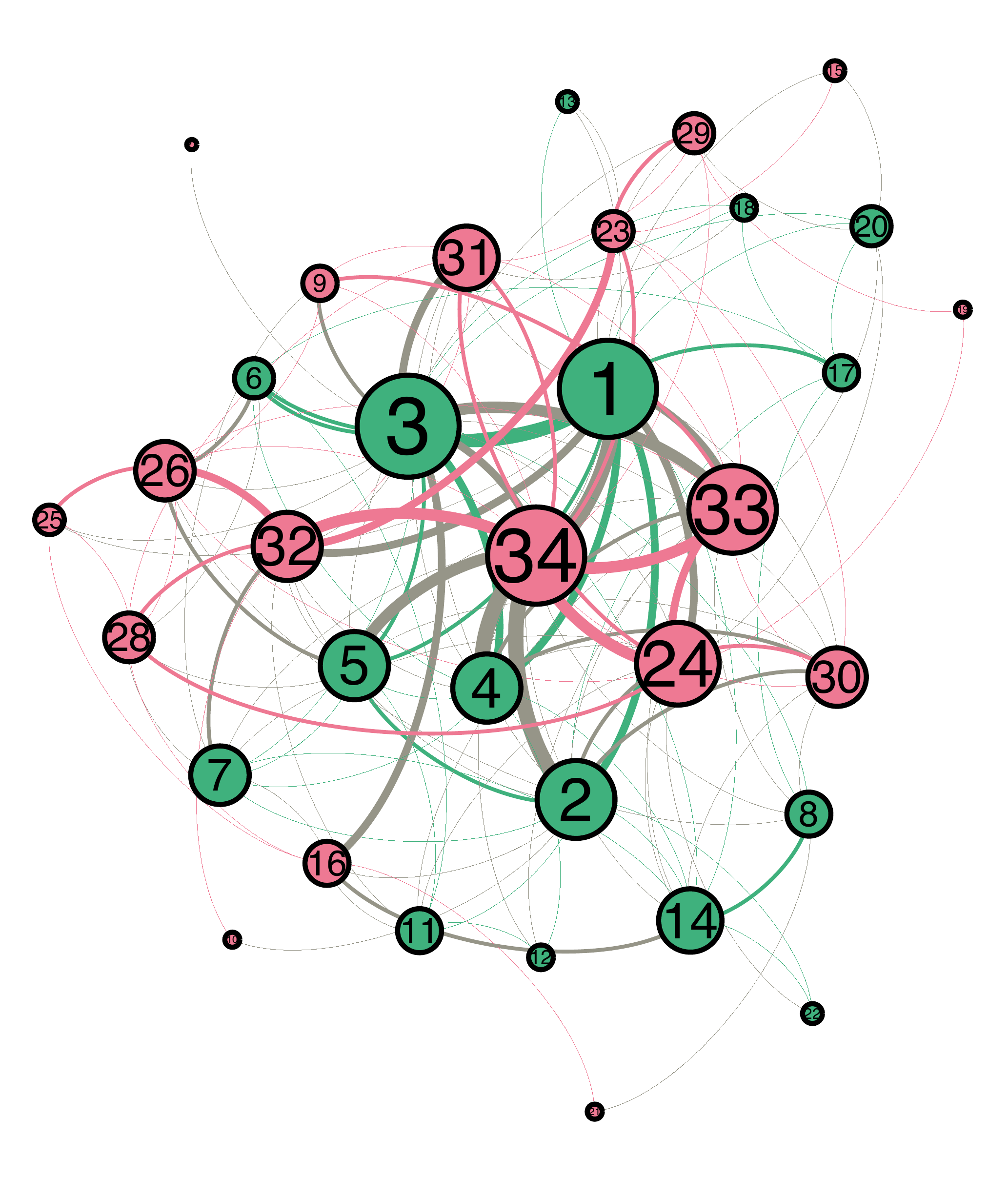}}}\hfill
  \subfigure[\label{fig:karate:biased}]{\fcolorbox{biasnet}{biasnet}{\includegraphics[width=.3\columnwidth]{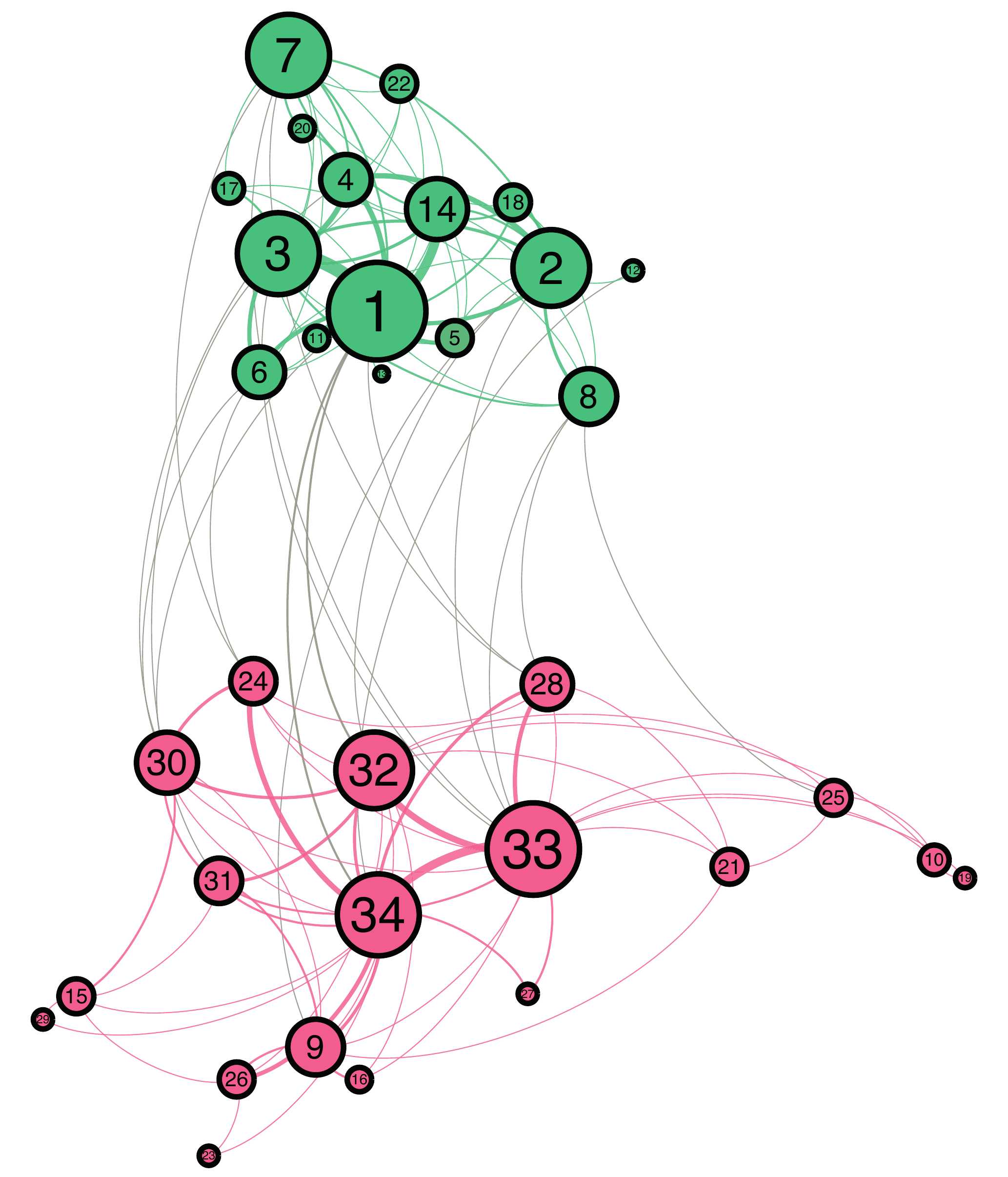}}}\\
  \subfigure[\label{fig:karate:histogram}]{\includegraphics[width=\columnwidth]{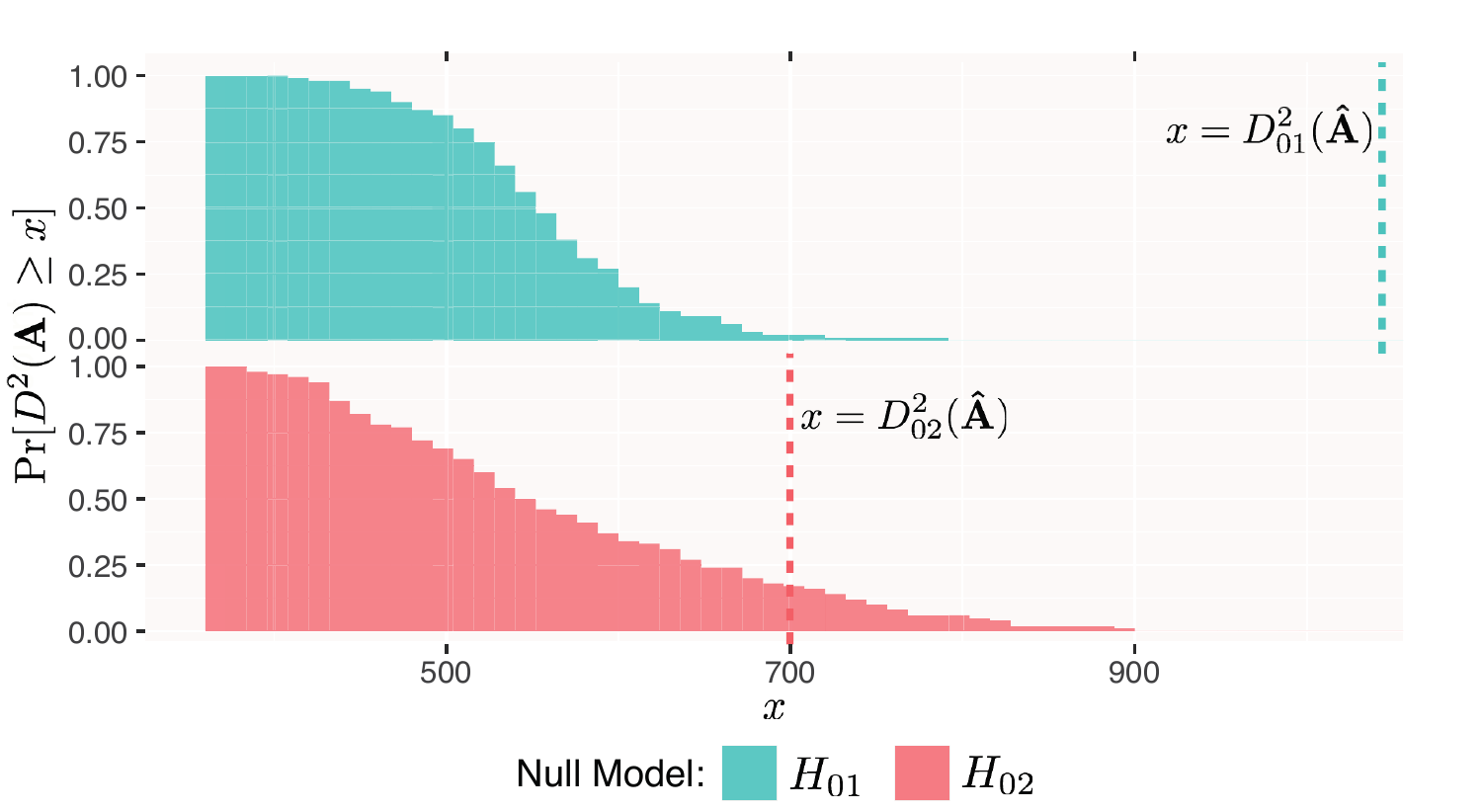}}\caption{(a) shows the empirical Karate club network, (b) shows a random realization drawn from the (unbiased) hypergeometric ensemble (cf. Eq.~\ref{eq:hypergeometricNet}), while (c) shows a random realization drawn from a generalized hypergeometric ensemble with a block matrix $\mathbf{\Omega}$ (cf. Eq.~\ref{eq:walleniusNet}). The top panel of (d) shows the ccdf of Mahalanobis distances obtained for $5000$ random realizations of the (unbiased) hypergeometric ensemble. The bottom panel of (d) shows the CCDF of Mahalanobis distances for the generalized hypergeometric ensemble with block matrix $\mathbf{\Omega}$. Dashed lines indicate the Mahalanobis distance for the observed network in the two ensembles.}
\label{fig:karate}
\end{figure}

We illustrate the testing procedure using Zachary's Karate Club network, denoted as $\mathbf{\hat{A}}$, with a well-known community structure~\cite{Zachary} shown in Fig.~\ref{fig:karate:empirical}.
Our first (null) hypothesis $H_{01}$ is that the network contains no patterns that cannot be explained by its degree sequence.
This hypothesis can be encoded by means of $\mathbf\Omega_{01} \equiv 1$, which corresponds to the configuration model.
We test this hypothesis by computing $\Pr\left[D_{01}^{2}(\mathbf{A})\geq D_{01}^{2}(\mathbf{\hat{A}})\right]$ based on the distribution of Mahalanobis distances for random realizations drawn from the ensemble (see top panel of Fig.~\ref{fig:karate:histogram}).
As expected, we obtain $p \approx 0$, i.e. we can safely reject hypothesis $H_{01}$.
This can be intuitively confirmed by visually comparing the empirical network in Fig.~\ref{fig:karate:empirical} to a random realization of the (unbiased) ensemble with $\mathbf\Omega_{01} \equiv 1$ shown in Fig.~\ref{fig:karate:unbiased}.

As the second (null) hypothesis $H_{02}$ we consider that the network topology is solely explained by the presence of two communities, where pairs of nodes within a community have higher propensities than nodes in different ones.
Similar to a stochastic block model~\cite{Peixoto2014}, this hypothesis can be encoded by a simple block matrix structure, where we set $\Omega_{ij} = 1$ for all pairs $i,j$ in the same community, while $\Omega_{ij} = \alpha<1$ otherwise.
Choosing $\alpha$ as the observed fraction of edges across communities allows us to calculate the distribution of Mahalanobis distances for random realizations of the resulting statistical ensemble (cf. bottom panel of Fig.~\ref{fig:karate:histogram}).
From this, we obtain $p=0.158367$, which does not allow us to reject hypothesis $H_{02}$.
Again, this result can be intuitively confirmed by visually comparing the empirical Karate club network shown in Fig.~\ref{fig:karate:empirical} with the random realization generated from the block matrix model shown in Fig.~\ref{fig:karate:biased}.
The example shows that a generative model \emph{only} accounting for heterogeneous node degrees and community structure is sufficient to explain the observed network.
Moreover, this highlights how the known functional form of distribution, expected values and covariance provided by our ensemble formulation provides a novel approach to (i) statistically test hypotheses in networks, and (ii) assess the significance of community structures.

In conclusion, we introduced \emph{generalized hypergeometric ensembles}, a broad class of statistical ensembles that allows to encode a wide-range of topological patterns.
Unlike similar approaches, it provides analytical expressions for important statistical quantities like expected values and covariance.
Through this, the novel class of ensembles introduced in this Letter provides broad perspectives for the analysis of complex networks, with applications in pattern recognition, hypothesis testing and statistical inference.
It opens new paths for the analytical study of widely-used ensembles defined, e.g., by the configuration model, the stochastic block model or Exponential Random Graph Models.
Our work thus contributes to the fundamentals of network analysis, with applications in the interdisciplinary study of complex systems in physics, biology, and (computational) social science.

\begin{acknowledgments}
We thank Pavlin Mavrodiev for discussions.
V.\,N., I.\,S.\ and F.\,S.\ acknowledge support from the Swiss National Science Foundation (SNF) through grant  CR31I1\_140644.
IS acknowledges support from the Swiss State Secretariat for Education, Research and Innovation (SBFI), Grant No.\ C14.0036 and the European Cooperation in Science and Technology (COST) project TD1210.
\end{acknowledgments}

\bibliography{bibliography}

\end{document}